\documentstyle[12pt,bezier,emlines]{article}
\title{\bf Intrinsic mechanism of phase locking in two-dimensional 
  Josephson junction networks in presence of an external magnetic
field}
\author{ W. Krech\thanks{owk@rz.uni-jena.de}\hskip4mm and
K. Yu. Platov\thanks{okp@rz.uni-jena.de}\\
\sl Friedrich-Schiller-Universit\"at Jena\\
\sl Physikalisch-Astronomisch-Technikwissenschaftliche Fakult\"at\\
\sl Institut f\"ur Festk\"orperphysik\\
\sl 07743 Jena\\
\sl Max-Wien-Platz 1}
\begin{document}
\maketitle
\begin{abstract}

   We present  numerical simulations of the dynamics of 
two-dimensional Josephson junction arrays to study the 
mechanism of mutual phase locking. 
We show that in the 
presence  of an external magnetic field two mechanisms 
are playing a role in phase locking: feedback through 
the external load and internal coupling between rows due
to microwave currents induced by the field. 
We have found the parameter values (junction capacitance, 
cell loop inductance, impedance of 
the external load) for which the interplay of both these mechanisms
leads to the in-phase solution. 
The case of unshunted arrays is discussed as well.


\end{abstract}
\vskip1cm
\centerline{Subm. to {\it Europhysics Letters}}

\vfill
\eject
%\section{Introduction}

 Josephson junctions are natural voltage-controlled 
oscillators [1].
By building up one-dimensional (1D) and 
two-dimensional (2D) arrays of the junctions one can hope 
to obtain the necessary output power which is required for 
many practical applications. This goal will be achieved 
when all junctions of the array are phase-locked. 
While  a detailed theoretical analysis of phase locking in 1D 
systems has been performed several years ago [2],[3] 
and has been supported
by promising recent experiments [4]
the  full examination of the problems connected with the formation of  
stable phase locking in 2D arrays  has not been achieved up to now. 
Several authors have shown that 2D arrays are more stable 
against non-uniformities in the critical currents [5]-[8].
Nevertheless, it is not understood so far  whether  2D 
arrays possess an internal mechanism of coupling which could be
a potential advantage. Even a qualitative 
observation regarding the mechanism could provide useful 
criteria for the problem of optimum array design.

 In this letter we address these issues by means of simulations 
of the dynamical properties of 2D Josephson networks with and 
without external load in presence of an external magnetic 
field.   

%\section{Basic structure and equitions}
   
We investigate  arrays consisting of active junctions connected by 
superconducting inductances in the direction perpendicular to the 
bias current.
Such a structure has been accepted as a standard design now
 [8]-[10]. We have $N$ rows of $M$ junctions.
The behavior of every Josephson junction is simulated  
using the RSJ model. Within this model the current through the $k$th 
junction is given by 
\vskip1mm
\begin{equation}
 \beta_c\ddot{\varphi_k}+\dot{\varphi_k}+\sin\varphi_k={i_k},
\end{equation}
\centerline{$i_k=I_k/I_C$.}
\vskip5mm
Here $\beta_c=2\pi CI_C{R_N}^2/\Phi_0$ is the McCumber parameter, 
$\varphi_k$ is the Josephson phase difference  over the 
junction, $\Phi_0$ stands for the magnetic  flux quantum.
  $C$, $R_N$ and $I_C$ are the junction capacitance, 
normal resistance and critical current, respectively.
  $I_k$ is the total current flowing through the $k$th junction.
Introducing  dimensionless parameters, we can formulate  flux quantization 
conditions for every cell  containing junctions {\it m} and {\it n}:
\vskip1mm
\begin{equation}
\varphi_m-\varphi_n={{\varphi}_{ext}}^{(m,n)}+\l_0 \sum_{p=1}^4 \mu^p 
i_p^{(m,n)},
\end{equation} 
\begin{equation}
\sum_{p=1}^4 \mu^p=1,
\end{equation}
where we introduced the  parameters
$ l_0=2 \pi I_c L_0 / \Phi_0$ and
${{\varphi}_{ext}}^{(m,n)}=2 \pi{{\Phi}_{ext}}^{(m,n)}/\Phi_0$. 
Furthermore, $ i_p^{(m,n)}= I_p^{(m,n)}/I_c$ is the normalized
current flowing 
through the respective inductance of the {\it p}th branch. 
The quantity  $\mu^p$ is the contribution to the
loop inductance $l_0$ from the  {\it p}th branch  of the cell.
The frequency of the Josephson oscillations is 
\begin{equation}
 \omega_j = \overline{v}_j;
\end{equation}
here $\overline{v}_j$ indicates the normalized DC component 
of the voltage across junction $j$.
We restrict our simulations to the case
of a uniform external magnetic flux  ${{\varphi}_{ext}}^{(m,n)}= 
\varphi_{ext}$.
The resistor in the external load $r_s$ has been choosen to match 
the array normal resistance
$r_{array}=N r_N/M$.
The first and second harmonics of the 
current flowing through the external load have been calculated as
\begin{equation}
 i_{n\omega}=1/T \int^T_0 i(t)\exp{(- in \omega t)} \, dt  , \: \: n=1,2
\end{equation} 
with the averaging time $T\ge 1000/ \omega $.
In the simulations we also  introduced a small spread of 
the critical currents $\sigma $ of about 5 \%
with the product $V_c=I_cR_N$ beeing the same for  
all junctions.
The arrays have been investigated  by means of the 
PSCAN  program [11],[12]. 

   The most effective phase
locking mechanism for both 1D and 2D arrays is high-frequency coupling
between an array and external load,  as has been discussed 
by several authors  before [1].
	In dependence on inductive or
capacitive character of the load one can expect 
different types of coherent solutions. 
The in-phase state  of a 1D array is characterized by conditions 
\begin{equation}
\varphi_i - \varphi_j = 0   
\end{equation} 
for the phase difference between any two junctions. 
This and other types of coherent solutions 
for the 1D have been described 
in the papers [3],[13].  For the 2D array
with $\varphi_{ext}=0$ and all junctions beeing identical
we can write down the same relations for any phases in both 
"transverse" (first index) and "longitudial" (second index)
directions
\begin{equation}
 \varphi_{i,k} - \varphi_{j,l} = 0 . 
\end{equation}
 
  The situation changes when  we consider 2D structures
with nonzero magnetic field  within the cells. In this case  
the array symmetry is broken and we can no longer expect
uniform oscillations of all junctions within a  row.
However, we still can define  an in-phase solution
\begin{equation}
\varphi_{i,k} - \varphi_{i,l} = 0 . 
\end{equation}
with respect to the "longitudial" direction.
In the following we have to  find conditions under which 
an external magnetic flux does not disturb this 
in-phase solution. We show
that the 2D array {\it standard design itself } can not 
assure this solution and must be improved. 
For the sake of clarity we start with the four junction array 
(Fig. 1) and  the traditional design of the unit cell. 
In this case the additional branch {\bf A}
is absent. Only the external load {\bf B} must control the
phase locking of the whole system. 

   In Fig. 2 we show the results of simulations which have 
been performed for the value  $\l_0=0.5$ of the loop inductance
and two different types of external load. For the capacitive load 
(solid line in Fig. 2a) we have obtained the anti-phase solution
within the whole range of the external magnetic field. 
We find no first harmonic 
in the  net oscillation. Consequently, the microwave current
through the load  is characterized  by
the  second harmonic component  essentially (Fig. 2b). 
The phase relations between 
the junctions in each column remain the same
for all values of the external field. On the other hand, in 
the inductive load regime (filled 
triangles in Fig. 2a) 
we have got in-phase oscillations within 
the regions $\varphi_{ext} < 2.0$ and $\varphi_{ext}> 4.2$
and the anti-phase solution for $ 2.0 < \varphi_{ext} < 4.2.$ 
It is worth noting that an increase of the magnetic flux
 ($0 \le \varphi_{ext} \le \pi $) leads  to  
an increasing oscillation frequency. As a result,
the transition observed can not be caused by the action
of the external load which remains  in the inductive  regime.

  In order to explain the result  we turn to 
the properties of the unshunted array first.
Here, the microwave currents induced by external magnetic fields 
can contribute not only to  the phase locking within one row 
but also to the phase locking between neighbouring rows. 
A stability analysis  for zero Josephson junction 
capacitance $\beta_c=0$ and $l_0 \ll 1 $ 
has been performed analytically [14]. 
As a result one observes that  even small magnetic
fields favour the stability of the anti-phase state.

  For  non-zero capacitance of the Josephson junctions and 
for the unit cell inductance  $l_0 \approx 1$  the 
frequency dependence of both in- and anti-phase states is more 
complex. The boundaries of the stability regions can be determined
numerically only. In Fig. 3 we show these regions 
for  the network without external load
and  $\varphi_{ext}=1.5$. One recovers that the point 
 $l_0=0.5$ and $\beta_c=0.0$ 
(which are our parameters for Fig. 2)  lies within the region 
where the unshunted array exhibits a stable anti-phase solution.

  The state choosen by the array  {\it with shunt} depends 
on the interplay of the two high-frequency currents, flowing through
a given junction of the array, i.e.  the shunt current 
and the circulating current.
For a small value of the 
external field the shunt current is strong
enough to support the in-phase state. The increasing magnetic 
field within the cells produces a shift of the oscillation 
phases along the vertical 
direction and  leads to a decreasing  of the microwave current through 
the external load (Fig. 2b). The mechanism providing 
the in-phase solution 
weakens and for some value  $\varphi_{ext}^{lim}$ fails.  

   Our simulations for the shunted array with a larger number of  
junctions (up to 6 in every row) have shown the following:
The value of the  critical flux $\varphi_{ext}^{lim}$ 
for which switching into anti-phase
state takes place becomes smaller with the number $M$ of 
Josephson junctions per row growing. From a practical point 
of view this means that the 
increasing of the array size makes the in-phase state less tolerant 
toward the external field. 
For a 4 $\times$ 4 shunted array with the loop inductance $l_0=1.0$
this limiting value is $\varphi_{ext}^{lim}=0.7$. After switching the 
array reveals a different behavior: The phase differences of
the junction voltage oscillations are equal to
 $\pi$ for nearest neighbours in each column. 
Large microwave  currents are flowing along the transverse inductances. 

   Another conclusion we have made from the simulations of the arrays
with $M,N \ge 2$ is
that the boundaries of stability of the in-phase 
and anti-phase solutions for the unshunted array
at the  $l_0, \beta_c$-plane do not depend on $M$ and $N$. 
One possible explanation is that the circulating microwave 
currents choose the nearest ways to flow and consequently only 
the short-range coupling between neighbouring cells
determines the stability conditions.

The mechanism assuring the in-phase state
depends mainly on the impedance in the vertical direction 
(in contrast to the other one created by circulating currents).
This way one can hope to influence the relative strenght
of both mechanisms 
by changing the impedance of the cell into the direction 
perpendicular to the bias current.  
   We have done so by including the additional branch {\bf  A}.
Fig. 4 presents the boundaries of the in-phase and anti-phase solutions
for the unshunted (stars) and shunted (filled boxes) arrays 
in comparison. The additional branch {\bf A} improves the 
stability of the in-phase state within the
region $l_0 \le 1.0, \beta_c \le 1.0 $ for the unshunted array (this
set of parameters is supposed to be the most promising for obtaining  
maximum output power). As a result,  the shunted array remains phase locked 
for all values of the external flux (Fig. 5).

  In conclusion, we have shown that the magnetic field in two-dimensional
arrays provides an internal mechanism  of phase locking.
This mechanism yields the stability of the anti-phase state for the array
with inductances only beeing used for the transverse connections.
A capacitive shunt parallel to these inductances improves the stability
of the in-phase state. 

The authors  would like to express their 
thanks to Deutsche Forschungsgemeinschaft for supporting this work. 
\eject

\eject
\centerline{FIGURE CAPTIONS}
\vskip10mm
% Fig. a951a.lp
{
Fig. 1. An array of 2$\times$2 Josephson 
junctions with an additional  branch  $r_x,c_x$ (denoted by {\bf A})
and external shunt   $r_s,c_s,l_s$ ({\bf B}); $l_x=(1/4)l_0, \: 
l_y=(1/4)l_0$.

}
\vskip10mm

%Fig flux1.t
{
 Fig. 2. (a) The frequency $\omega $ of a shunted 
array with inductive (filled triangles) and 
capacitive (solid line) shunt as a
function of the external flux $\varphi_{ext}$ 
(Parameters: 
$i_{dc}=1.5,\: \beta_c=0.0,\: l_0=0.5, \: l_s=1.0, \: r_s=1.0$).
}{
(b) The first and second harmonics of the shunt
current $i_{\omega}$ (solid line), $i_{2\omega}$ ( dot line) 
for the inductive shunt ($c_s=2.0$) and the second harmonic (open triangles)
of the shunt current for the capacitive shunt ($c_s=0.2$) as a
function of the external flux $\varphi_{ext}$ 
(Parameters: $i_{dc}=1.5, \: \beta_c=0.0, 
\: l_0=0.5, \: l_s=1.0, \: r_s=1.0$).
}
\vskip10mm
% Fig.  bound2.t 
{
 Fig. 3. The boundary of the in-phase
and  anti-phase solutions at the $\beta_c, \l_0$-plane 
for the unshunted array  from Fig. 1 
(Parameters: $i_{dc}=1.5, \: \varphi_{ext}=1.5$).
}
\vskip10mm
% Fig.  bound1.t 
{
 Fig. 4. The boundary of the  in-phase
and  anti-phase solutions at the  
$\beta_c,\l_0$-plane for unshunted array (stars) 
and shunted array (filled boxes) with the 
additional branch {\bf A}
(Parameters: $i_{dc}=1.5, \: \varphi_{ext}=1.5, 
\: l_s=1.0, \: r_s=1.0, \: c_s=2.0, \: c_x=6.0, \: r_x=0.1$). 
}
\vskip5mm
%Fig flux2.t
{
 Fig. 5. The frequency $\omega $ of the shunted 
array without (stars) and 
with (solid line) additional branch {\bf A} as a
function of the external flux $\varphi_{ext}$ 
(Parameters: 
$i_{dc}=1.5, \: \beta_c=0.0, \: l_0=0.5, \: l_s=1.0, \: r_s=1.0$).
}
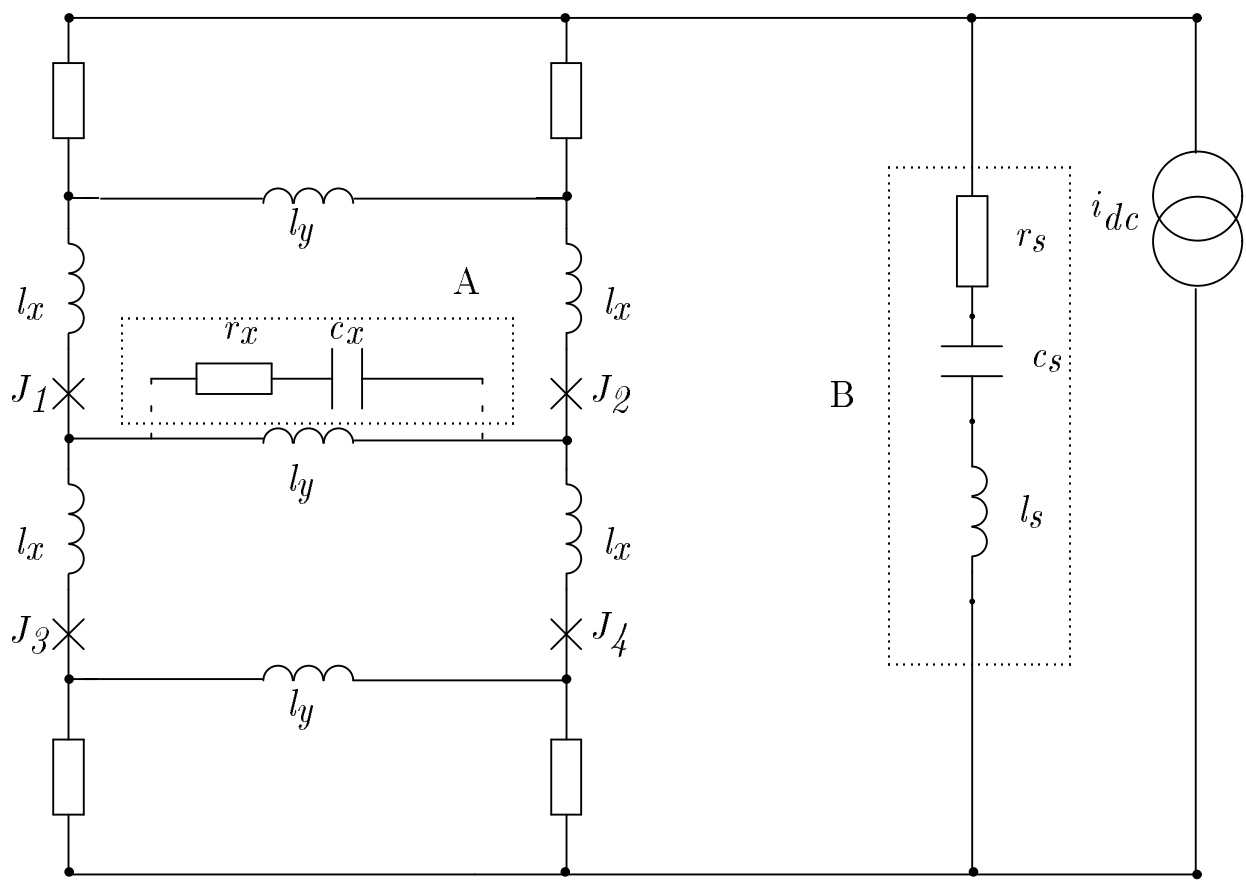
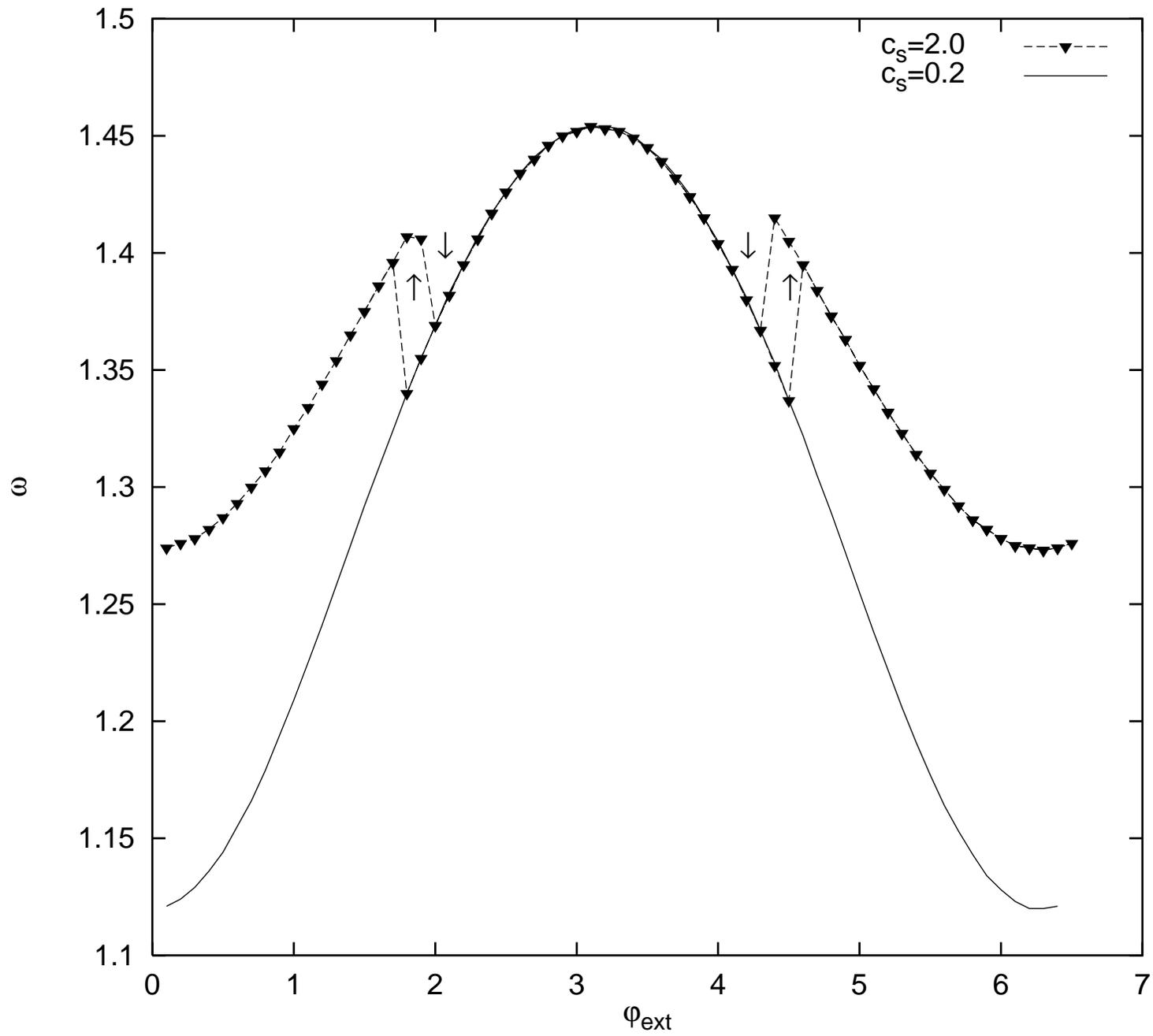
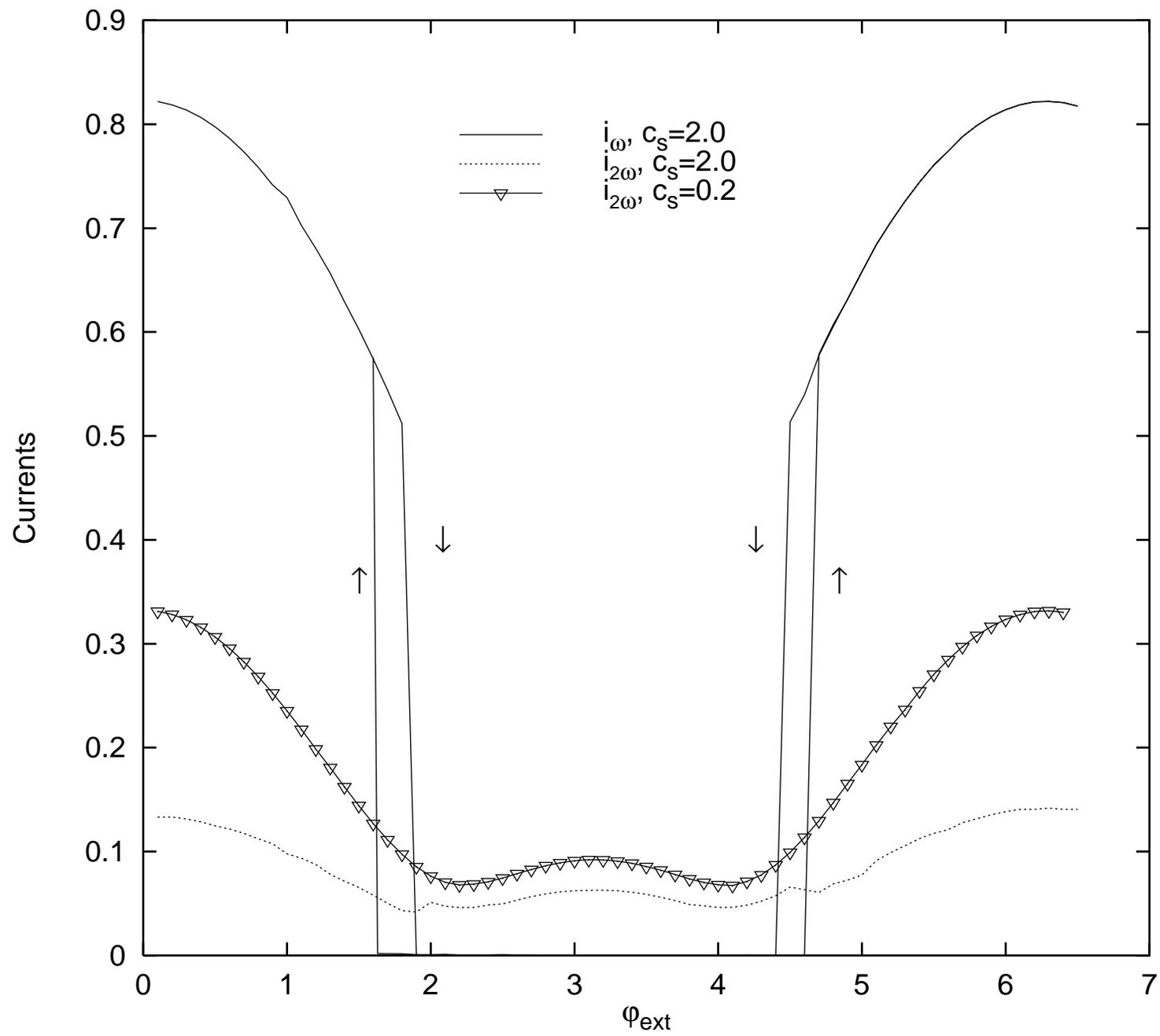
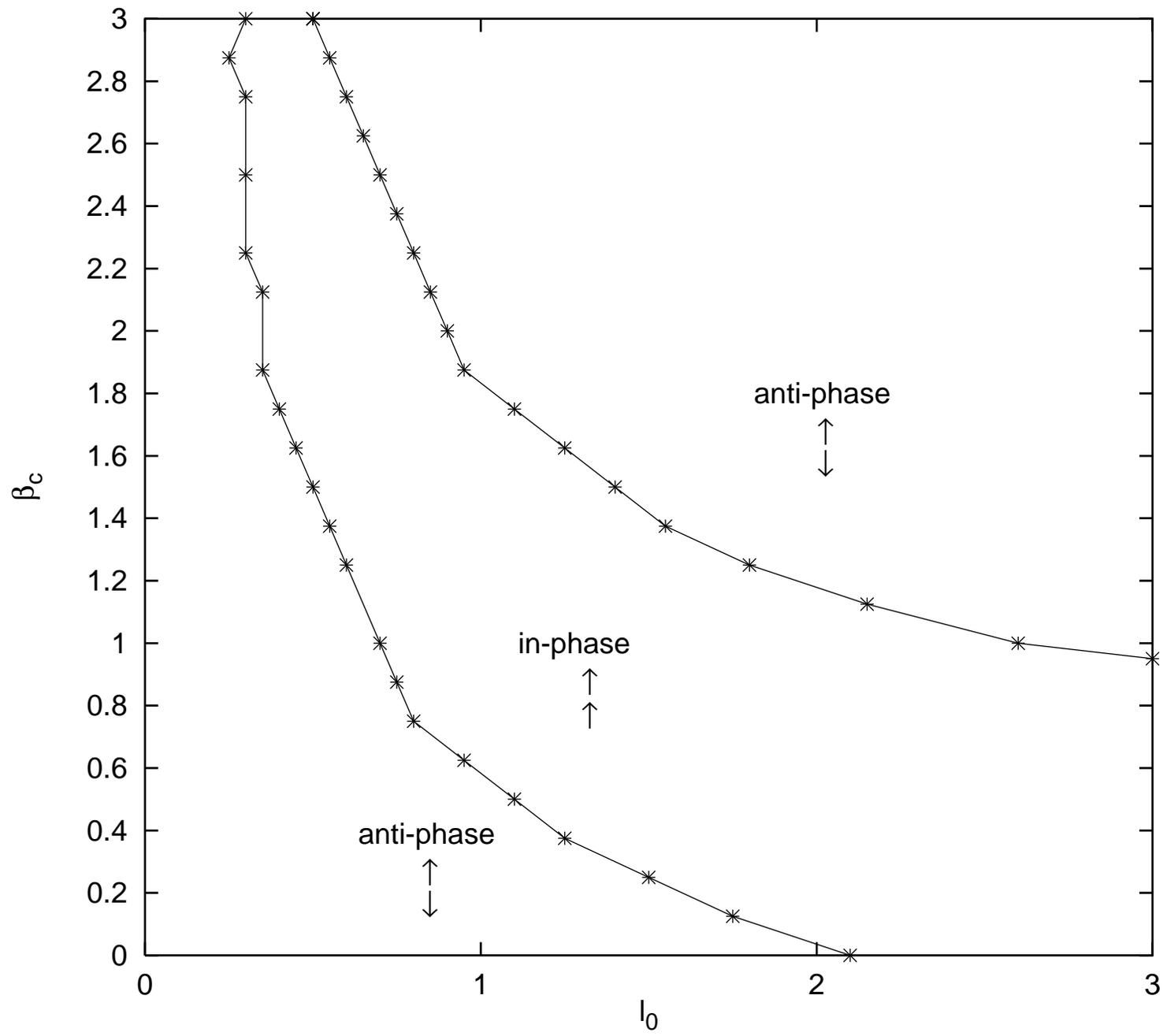
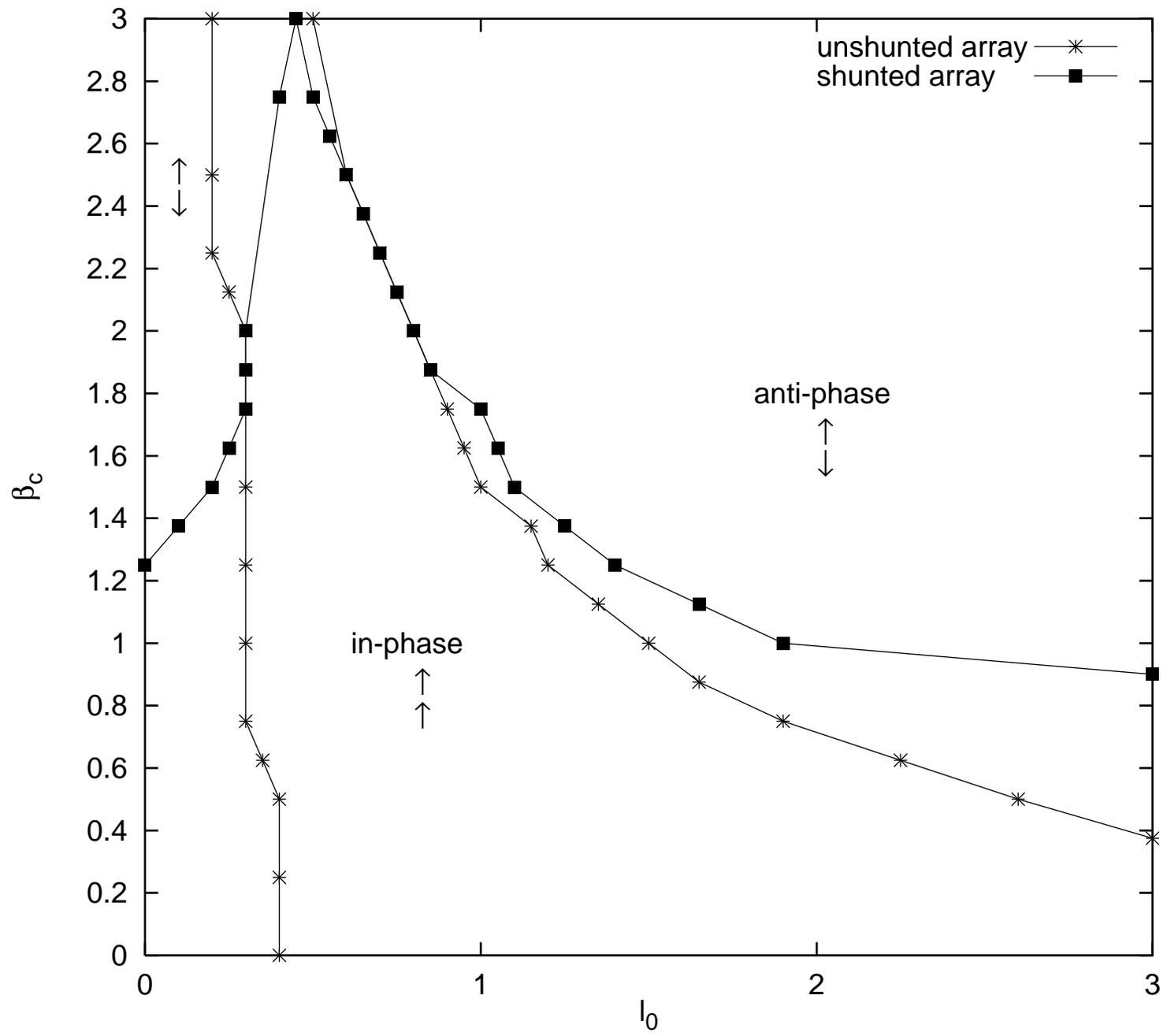
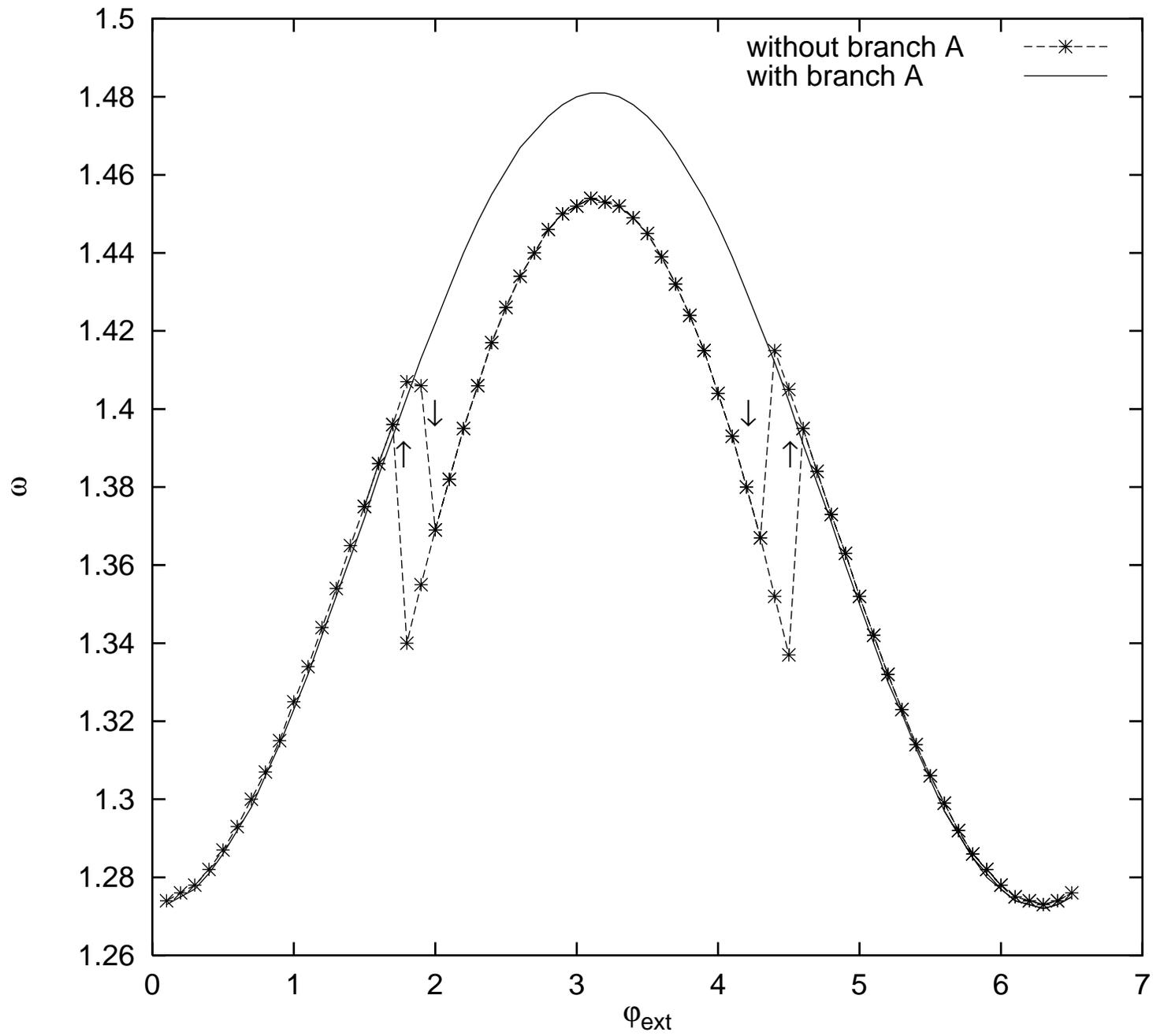
\end{document}